\documentclass[graybox]{svmult}
\usepackage{geometry}                
\usepackage{graphicx}
\usepackage{amssymb}
\usepackage{url}
\usepackage{subfigure}
\usepackage{makeidx}

\makeindex

\begin{document}

\title*{Physical maze solvers. \\All twelve prototypes implement 1961 Lee algorithm.}

\author{Andrew Adamatzky}

\institute{Andrew Adamatzky \at University of the West of England, Bristol, UK \email{andrew.adamatzky@uwe.ac.uk}}

\maketitle




\abstract{We overview experimental laboratory prototypes of maze solvers. We speculate that all maze solvers 
implement Lee algorithm by first developing a gradient of values showing a distance from any site of the maze to the destination site and then tracing a path from a given source site to the destination site. All prototypes approximate a set of  many-source-one-destination paths using resistance, chemical and temporal gradients. They trace a path from a given
source site to the destination site using electrical current, fluidic, growth of slime mould, Marangoni flow, crawling of epithelial cells, excitation waves in chemical medium, propagating crystallisation patterns. Some of the prototypes 
visualise the path using a stream of dye, thermal camera or glow discharge; others require a computer to extract
the path from time lapse images of the tracing. We discuss the prototypes in terms of speed, costs and durability of the path visualisation. 
}

\tableofcontents

\markboth{Adamatzky: Physical Maze Solvers}{Adamatzky: Physical Maze Solvers}

\section{Introduction}

To solve a maze is to find a route from the source site to the destination site.  
If there is just  a single path to the destination the maze is called a labyrinth. \index{maze} \index{labyrinth}  To solve a labyrinth one must just avoid dead ends. In a maze there are at least two paths leading from the entrance to the exit. To solve a maze one must find a shortest path. Not rarely concepts of `maze', `labyrinth' and `collision-free shortest path' are mixed in experimental laboratory papers. We will not differentiate either.  All algorithms and physical prototypes that solve shortest collision-free path on a planar graph solve mazes~\cite{blum1978power}. All algorithms and prototypes that solve mazes solve labyrinths. 

There are two scenarios of the maze problem: the solver does not know the whole structure of the maze 
and the solver knows the structure of the maze.

The first scenario --- we are inside  the maze --  is the original one. This is how Theseus, \index{Theseus} the Shannon's maze solving mechanical mouse,  was born~\cite{shannon1951presentation}.\footnote{See \url{http://cyberneticzoo.com/mazesolvers/}} The mouse \emph{per se} was a magnet with copper whiskers. The mechanism was hidden under the maze. A circuit with hundred of relays and mechanical drives grabbed the mouse from below the floor and moved to a randomly chosen direction. When the mouse detected an obstacle with its whiskers the underfloor mechanism moved the mouse away from the obstacle and other direction of movement has been selected. The task was complemented when the destination site was found. Being placed at any site of the maze the mouse was able to find a path towards the destination site. Several electro-mechanical devices have been built in 1950-1970, including well know Wallace's maze solving computer~\cite{wallace1952maze}. The Theseus also inspired a range of robotic mice \index{robotic mouse} competitions~\cite{zhang2014study}. The algorithms of a maze traversing agent, see overviews in ~\cite{willardson2001analysis, babula2009simulated} include  random walk~\cite{aleliunas1979random}; the Dead Reckoning (the mouse travels straight, when it encounters a junction it turns randomly, when it finds itself in the dead end it turns around), Dead End Learning (the agent remembers dead ends and places a virtual wall in the corridor leading to each dead end); Flood Fill (the agent assigns a distance, as crow flies, to each site of the maze and then travel in the maze and updates the distance values with realistic numbers); and, Pledge algorithm where the maze traversing agent is equipped with a compass, which allows to maintain a predetermined direction of motion (e.g. always north); the intersections between the corridors oriented north-south and  walls are treated as graph vertices~\cite{blum1978power, abelson1986turtle}.  Hybrids of the Wall Follow and the Pledge algorithms are used in industrial robotics and space explorations: a robot knows coordinates of the destination site, has a compass, turns on a fixed angle and counts turns~\cite{lumelsky1987path, lumelsky1991comparative}.There are also  genetic programming and artificial neural networks for maze solving~\cite{khan2010solving}.

The second scenario of maze solving --- we are above the maze --- is the one we study here.  In 1961 Lee proposed the algorithm~\cite{lee1961algorithm, rubin1974lee} \index{Lee algorithm} which became one of the most  famous, reused and rediscovered algorithms in last century. We start at the destination site. We label neighbours of the site with `1'. Then we label their neighbours with `2'.  Being at the site labelled $i$ we label its non-yet-labelled neighbours with $i+1$.  Sites occupied by obstacles, or the maze walls, are not labelled. When all accessible sites are labelled the exploration task is completed. 
To extract the path from any given site of the maze to the destination site we start at the source site. Then we select 
a neighbour of the source site with lowest value of its label. We add this neighbour to the list.  We jump at this neighbour. Then we select its neighbour with lowest value of the label. We add this neighbour to the list. We jump at this neighbour. We continue like that till we get at the destination site. Thus the algorithm computes one-destination-many-sources shortest path.  A set of shortest path starting from each site of the maze gives us a spanning tree which nodes are sites of the maze and a root is the site where wave pattern of labelling started to grow. In robotics the Lee algorithm was transformed into a potential method \index{potential method} pioneered in  \cite{pavlov1984method} and further developed in \cite{wyard1995potential, hwang1992potential}. The destination is assigned an infinite potential. Gradient is calculated locally. Streamlines from the source site to the destination site are calculated locally at each site by selecting locally maximum gradient~\cite{connolly1990path}. Also, some algorithms assume that the destination has an `attracting' potential and obstacles are `repellents'~\cite{khatib1986real, adamatzky2002collision, adamatzky2006utopian}. 

Most experimental laboratory prototype of maze solvers implement the Lee algorithm. The gradients developed are resistance (Sect.~\ref{resistance}), chemical (Sect.~\ref{resistance}), temporal (Sect.~\ref{resistance}) and thermal (Sect.~\ref{resistance}).  The paths are traced along the gradients by electrical current (Sect. \ref{electrical_current}),
fluids (Sect.~\ref{fluids}), cellular cytoplasm (Sect.~\ref{cytoplasm}), Marangoni flow (Sect.~\ref{marangoniflow}),  living cells (Sect.~\ref{livingcells}), excitation waves (Sect.~\ref{excitationwaves}) and crystallisation (Sect.~\ref{crystallisation}). We present brief descriptions of known experimental laboratory prototypes of maze solvers 
and analyse them comparatively.

\section{Resistance gradient}
\label{resistance}

Imagine a maze filled with hard balls. The entrance and the exist are open. We put our hand in the entrance and push the balls. Balls in the dead ends have nowhere to move. The pressure is eventually transferred to the balls nearby exit. 
These balls start falling out.  We add more balls thought entrance and push again. Balls fall out through the exit. Thus 
a movement of balls is established. The balls are moving along the shortest path between the entrance and the exit. 
The balls explore the maze in parallel and `calculate' the path from the exit to the entrance. In this section we discuss prototypes which employe electrical and hydrodynamic resistances.

\subsection{Electrical current}
\label{electrical_current}

Approximation of a collision-free path \index{collision-free path} with a network of resistors \index{resistor} is proposed in \cite{tarassenko1991analogue, tarassenko1991real}. A space is discretised as a resistor network. The resistors representing obstacles are insulators or current thinks. \index{electrical current} Other resistors  have the same initial resistance. An electoral power source is connected to the destination and the source sites.  The destination site is the electrical current source~\cite{bugmann1995route}. 
Current flows in the grid. Current does not flow into obstacles. To trace the path one must follow a current streamline by performing gradient descent in electrical potential.  That is for each node a next move is selected by measuring the voltage difference between current node and each of its neighbours, and moving to the neighbours which shows maximum voltage. We are not aware of any large-scale prototype of such path solver.  Two VLSI processors \index{VLSI processor} have been manufactured~\cite{stan1994analog}. They feature  16$\times$16 and 18$\times$18 cells, 
2~$\mu$m nwell SCMOS technology, 3960~$\mu$ $\times$ 4240`$\mu$ and 4560~$\mu$$\times$4560~$\mu$ frame size, 16-bit asynchronous data bus. For gradient descent  the source is 5~V and the destination is 0~V, and \emph{vice verse} for gradient ascent. 

There are two ways to represent walls of a maze~\cite{bugmann1995route}: Dirichlet boundary \index{boundary condition}
conditions and Neumann boundary conditions.  When the Dirichlet boundary conditions is adopted the walls, or obstacles, 
have zero electrical potential and act as sinks of the electrical current~\cite{connolly1990path}. Then the current lines
are perpendicular to the walls and an agent, e.g. a robot, travelling along the current lines stays away from the walls~\cite{bugmann1995route}.  In case of the  Neumann boundary conditions the walls are insulators~\cite{tarassenko1991analogue}.  The walls are not `felt' by the electrical current. Then the current lines are parallel to the walls. This gives the travelling agent less clearance. The original approach of \cite{tarassenko1991analogue} has been extended to networks of memristors (resistors with memory)~\cite{pershin2011solving}, which could allow for computation of a path in directed planar graph.

A shortest path can be visualised, though not digitally recorded, without discretisation of the space.
A maze is filled with a continuous conductive material. Corridors are conductors, walls are insulators.
An electrical potential difference is applied between the source and the destination sites. The electrical current `explores' all possible pathways in the maze. As proposed 
in~\cite{ayrinhac2014electric} the electrons, driven by the applied electrical field, move along the conductive corridors in a maze until they encounter dead end or the destination site. When the electrons reach dead end they are cancelled inside the conductor. The electric field inside the dead ends becomes zero.  The flow of electrons on the conductive pathways leading the destination does not stop.  Thus the electrical flow calculates the shortest path. This path is detected via glow-discharge or thermo-visualisation.

\subsubsection{{\sc \large glow}: Glow-discharge visualisation}
\label{glowdischarge}

This maze solver is proposed in \cite{reyes2002glow}. \index{glow-discharge}
A drawing of  a maze is transferred on a glass wafer and channels are etched in the glass.
 The channels are  c. 250~$\mu$m wide and  c. 100~$\mu$m deep. Electrodes are inserted in the source and the destination sites. The  glass maze is covered tightly and filled with helium \index{helium} at 500~Torr. A voltage of up to 30~kV, above the breakdown voltage, is applied. Luminescence \index{luminescence} of the discharge shows the shortest path in the maze. Shortest path is visualised in 500~ms. 

A maze solver using much less pressure and much lower voltage is proposed in \cite{dubinov2014glow}. 
The maze is made of a plexiglas disk, diameter 287~mm, 50~mm height. 
Channels 25~mm wide and~40 mm deep are cut in the disk. Hole for the anode is made in the center of the disk. Cathode is placed in the destination site. The maze is filled with air, pressure 0.110~Torr.
A gas-discharge chamber is   made of polyamide in 450~mm diameter and 50~mm height, copper cathode in the form of rectangular plate of 158~mm\textsuperscript{2} size is fitted on its side surface in a cathode holder.
The maze is placed in the gas-discharge chamber.  Stainless steel rod anode of 10~mm diameter is placed at the center of the chamber.   Voltage of 2kV is applied between the electrodes. The path is visualised by the glow of ionised air in the maze's channels. Experiments \cite{dubinov2014glow} also demonstrate propagation of striations, the ionisation waves~\cite{kolobov2013advances}, along the path.

\subsubsection{{\sc \large assembly}: Assembly of nano particles}

The maze is solved with nano particles in \cite{nair2015maze}. \index{nano particle} A maze is made of polydimethylsilicoxane and filled with silicon oil.  A drop of a dispersion of conductive nano particles: spherical copper nano particles 10~$\mu$m diameter and metallic carbon nanotubes 10~$\mu$m long --- is added at the source site. An electrical potential 1--5~kV is applied between the source and the destination.  The particles diffusing from the source site become polarised. The polarised particles experience dipole interactions. The dipole interactions make the particles to form a chain along line of the electric field with maximal strength. The chain is formed to maximise the electrical current and to minimise the potential drop.  The chain of the particles forms a conductive bridge between the electrodes at the source site and the destination site which represents the shortest path.

\subsubsection{{\sc \large thermo}: Thermo-visualisation}

A  maze solver using electrical current is proposed in ~\cite{ayrinhac2014electric}. The prototype 
employs thermal visualisation \index{thermal visualisation} of the electrical current. 
A maze 10$\times$10~cm is made from copper tracks on a printed board. Copper tracks represent corridors.  The electrical current 2.4~A is applied between the source and the destination sites. The flow of electrons heats the conductor, due to Joule heating. \index{Joule heating}  A local temperature of the conductor is proportional to intensity of the flow. 
In experiments~\cite{ayrinhac2014electric} the temperature along the shortest path increased by 
c.~10\textsuperscript{o}C. The heating is visualised with the infrared camera. The shortest path is represented by the brightest loci on the thermographic image.

\subsection{{\sc \large fluidic}: Fluidic}
\label{fluids}

In a fluidic maze solver developed in \cite{fuerstman2003solving} a maze is the network of micro-channels. The network is sealed. Only the source site (inlet) and the destination site (outlet) are open. The maze  is filled with a high-viscosity fluid. A low-viscosity coloured fluid  is pumped under pressure into the maze, via the inlet. Due to a pressure drop between the inlet and the outlet liquids start leaving the maze via the outlet. A velocity of fluid in a channel is inversely proportional to the length of the channel. High-viscosity fluid in the channels leading to dead ends prevents the coloured low-viscosity fluid from entering the channels.  There is no pressure drop between the inlet and any of the dead ends. Portions of  the `filler' liquid leave the maze. They are gradually displaced by the colour liquid. The colour liquid travels along maximum gradient of the pressure drop, \index{pressure drop} which is along a shortest path from the inlet to outlet. When the colour liquid fills in the path the viscosity along the path decreases. This leads to increase of the liquid velocity along the path.  The shortest path --- least hydrodynamic resistance \index{hydrodynamic resistance} path ---  from the inlet to the outlet is represented by channels filled with coloured fluid.

In the experiments \cite{fuerstman2003solving} channel width  varied from c. 90~$\mu$m to 200~$\mu$m, maze size c. 40$\times$50~mm. The maze is filed with ethanol-based solution of bromophenoll. A dark coloured solution of a food dye in mix of water and ethylene glycol is injected in the maze at a constant flow, velocity c 5--10~mm/sec. A drop of pressure between the inlet and the outlet is c. 0.75-2.25~Torr.  The channels along the shortest path become coloured.  The path is visualised in half-a-minute.

\subsection{{\sc \large physarum I}: Slime mould}
\label{cytoplasm}

The prototype based on reconfiguration of protoplasmic network of acellular slime mould \emph{Physarum polycephalum} \index{Physarum polycephalum} is proposed in~\cite{nakagaki2001path}. The slime mould is inoculated everywhere in a maze. The slime mould develops a network of protoplasmic tubes spanning all channels of the maze.  Oat flakes are placed in the source and the destination site. A tube lying along the shortest (or near shortest) path between sites with nutrients develop increased flow of cytoplasm. This tube becomes thicker. Tubes branching to sites without nutrients become smaller due to lack of cytoplasm flow. They eventually collapse. The sickest tube represents the path between the sources of nutrients, and therefore, the path between the source and the destination sites. The selection of the shortest protoplasmic tube is implemented via interaction of propagating bio-chemical, electric potential and contractile waves in the plasmodium's body, see mathematical model in~\cite{tero2006physarum}.

\section{Diffusion gradient}
\label{chemical}

A source of a diffusing substance is placed at the destination site.  After the substance propagates all over the maze a concentration of the substance develops.  The concentration gradient is steepest towards the source of the diffusion. Thus starting at any site of the maze and following the steepest gradient one can reach the source of the diffusion. The diffusing substance represents one-destination-many-sources shortest paths. To trace a shortest path from any site, we place a chemotactic agent at the site and record its movement towards the destination site.

\subsection{{\sc \large marangoni}: Marangoni flow}
\label{marangoniflow}

A diffusion gradient \index{diffusion gradient}  determines a surface tension gradient. \index{surface tension} A flow of liquid runs from the place of low surface tension to the place of high surface tension.  This flow transports droplets.  \index{droplet} A maze solver proposed in~\cite{lagzi2010maze} is as follows.  A maze is made of polydimethylsiloxane, size c. 16 $\times$ 16~mm, channels have width 1.4~mm, and walls are 1~mm high. The maze is filled with a solution of potassium hydroxide. A surfactant is added to reduce the liquid's surface tension. An agarose block soaked in a hydrochloric acid is placed at the destination site. In c. 40~s a pH gradient establishes in the maze. Then a 1~$\mu$L  droplet of a mineral oil or dichloromethane mixed 
with 2-hexyldecanoic acid is placed at the source site. The droplet does not mix with the liquid filling the maze. 
The  droplet moves along the steepest gradient of the potassium hydroxide. The steepest gradient is along a shortest path.  Exact mechanics of the droplet's motion is explained in~\cite{lagzi2010maze} as follows. Potassium hydroxide, which fills the maze, is a deprotonating agent. Molecule of the potassium hydroxide removes  protons from molecules of  2-hexyldecanoic acid diffusing from the droplet. A degree of protonation is proportional to concentration of hydrochloric acid, diffusing form the destination site. Protonated 2-hexyldecanoic acid at the liquid surface determines the surface tension. The gradient of the protonated acid determines a gradient of the surface tension.  The surface tension decreases towards the destination site. A flow of liquid --- the Marangoni flow \index{Marangoni flow} ---  is established from the site of low surface tension to the site of high surface tension, i.e. from the start to the destination site. The droplet is moved by the flow~\cite{lagzi2010maze}; see also discussion on mobility of surface in \cite{pimienta2014self} and more details on pH dependent motion of self-propelled droplets in \cite{ban2013ph}.

In the prototype ~\cite{lagzi2010maze} a path from the start site to the destination site is traced by a droplet but not visualised. To visualise the path fully one must record a trajectory of the droplet. A visualisation is implemented 
in~\cite{lovass2015maze}.  A dye powder, Phenol Rd, is placed at the start site. The Marangoni flow transports the dye form the start to the destination. The coloured channels represent a path connecting the source site and the destination.

Another prototype of a droplet maze solver is demonstrated in \cite{cejkova2014dynamics}. The maze 
c. 45$\times$75~mm in size, with channels c. 10~mm wide, is filled with water solution of a sodium decanoate. 
A nitrobenzene  droplet loaded with sodium chloride 
grains is placed at the destination site. A 5~$mu$L decanol droplet is placed at the source site. The sodium chloride diffuses from its host nitrobenzene droplet at the destination site. A gradient of salt is established. The gradient is steepest along a shortest path leading from any site of the maze to the destination site. 
A decanol droplet moves along the steepest gradient till  the droplet reaches the nitrobenzene with salt droplet at the destination site.

\subsection{Living cells}
\label{livingcells}

A source of a chemo-attractant \index{chemo-attractant} is placed at the destination site. The chemo-attractant diffuses along the channels of the maze. It reaches the destination site eventually. The maximum gradient is along the shortest path from any given site of a maze to the destination site. A living cell is placed at the source site.  The cell follows the maximum gradients thus moving along the shortest path towards the destination site.

\subsubsection{{\sc \large physarum II}: Slime mould}

The slime mould \index{slime mould} maze solver based on chemo-attraction is proposed in~\cite{adamatzky2012slime}. An oat flake is placed in the destination site. The slime mould \emph{Physarum polycephalum} \index{Physarum polycephalum} is inoculated in the source site. The oat flakes, or 
rather bacterias colonising the flake, release a chemoattractant. The chemo-attractant diffuses along the channels. 
The Physarum explores its vicinity by branching protoplasmic tubes into  openings of nearby channels.  When a wave-front of diffusing attractants reaches Physarum, the Physarum halts the lateral exploration. Instead it develops an active growing zone propagating along the gradient of the  attractant's diffusion. The problem is solved when Physarum reaches the source site. The sickest tube represents the shortest path between the destination site and the source site. 
 Not only nutrients can be placed at the destination site but  any volatile substances that attract the slime mould, e.g. roots of the medicinal plant \emph{Valeriana officinalis} \cite{ricigliano2015}.

\subsubsection{{\sc \large epithelium}: Epithelial cells}

Experimental maze solver with epithelial cells \index{epithelial cell} is proposed in \cite{scherber2012epithelial}.
Epithelial cells move towards sites with highest concentration of the epidermal growth factor (EGF). \index{epidermal growth factor}
An epithelial cell uptakes EGF.  Thus the cell depletes EGF's concentration in the cell's vicinity.
A 400~$\mu$m$\times$400~$\mu$m maze is made of orthogonal channels  c.  10~$\mu$m wide \cite{scherber2012epithelial}. The channels are filled with epithelium culture medium. 
There is a uniform distribution of the EGF inside the maze at the beginning of an experiment.  
The maze is placed in the medium with `unlimited' supply of  EGF.  
A cell is placed at the entrance channel. The cell enters the maze and crawls along its first channel.  
The cell consumes EGF and decreases EGF concentration in its own neighbourhood. 
EGF from all channels, accessible from the current position of the cell, diffuses towards the site with low concentration. Supply of EFG in channels ending with dead ends is limited. Unlimited supply of EGF into the maze is provided via exit channel. An EGF diffusion gradient from the exit through the maze to dead ends and the entrance is established. The cell follows the diffusion gradient. The gradient is maximum along the shortest path. The cell moves along the shortest path towards the exit. 

\section{Temperature gradient: {\sc \large temperature}}
\label{temperature}

A Marangoni flow \index{Marangoni flow} is a mass-transfer of a liquid from a region with low surface tension to a region with high surface tension~\cite{lovass2015maze}.  The mass transfer can move droplets, or any other objects, or dyes. Any methods of establishing a surface tension gradient is OK for tracing a shortest path with Marangoni flow.  
In \cite{lovass2015maze} a temperature gradient \index{temperature gradient } is used. A maze is made  from polydimethylsiloxane with channels 
1.4~mm wide and 1~mm deep. The maze is filled with hot, c. 99\textsuperscript{o}C, aqueous solution of sodium hydroxide with hexyldecanoic acid. A steel sphere, diameter 4~mm, is cooled with dry ice and placed at the destination site in the maze.  A phenol red dye powder is placed on the surface of the liquid at the start site. The cold sphere creates temperature gradient. The temperature gradient creates a surface tension \index{surface tension} gradient. The Marangoni flow is established along a shortest path from any site of the maze to the destination site.  The dye powder applied at the source site is dragged by the flow towards the destination site. The trace of the dye represents the shortest path.

\section{Temporal gradient}
\label{temporal}

A wave front advances for a fixed distance per unit of time in a direction normal to the front. 
A wave generated at the source site of a maze reaches the destination site along the shortest path.
The wave `finds' the exit. We just need to record the path of the wave.

\subsection{{\sc \large vlsi}: VLSI array processor}

An array processor \index{array processor} solving shortest path over terrain with elevations  is reported in~\cite{kemeny1992parallel}. This is a digital processor of 24 $\times$ 25 cells, manufactured with 2 $\mu$m CMOS. A cell size is  296 $\mu$m $\times$ 330 $\mu$m, the processor's size is 7.9~mm $\times$ 9.2~mm.  An elevations map is encoded to 255 levels of grey and loaded into the processor array. A signal is originated at the destination site. Wave-front of the signal propagates on the array. Each cell delays a signal by time proportional to the `elevation' value loaded into the cells. Then the cell broadcasts the value to its neighbours. When a processor receives signal, the incoming signal direction is stored and further inputs to the cell are ignored. Starting from each cell we can follow the fastest path towards the destination site.

\subsection{{\sc  \large wave}: Excitation waves }
\label{excitationwaves}

In~\cite{steinbock1995navigating} a labyrinth solution in excitable chemical medium \index{excitable medium} is proposed. 
A c. 3$\times$3~cm labyrinth is made of vinyl-acrylic membrane. The membrane is  saturated with 
Belousov-Zhabotinsky (BZ) mixture. \index{Belousov-Zhabotinsky reaction} Impenetrable walls are made by cutting away parts of the membrane. The channels are excitable. The walls are non-excitable. Excitation waves \index{excitation wave} are initiated at the source site by touching the membrane with a silver wire. The wave-fronts propagate with a speed of c. 2~mm/min. Dynamics of the waves is recorded with 50~sec intervals. The locations of the wave-fronts are colour-mapped: the colour depends on the time of recording. A shortest path from the source site to the destination site is extracted from the time lapse colour maps. In this setup excitation waves explore the labyrinth but the path is extracted by a computer.

The approach is slightly improved in \cite{agladze1997finding}.  By recording time lapse images of excitation 
wave fronts propagating in a two-dimensional medium we can construct  a set of isochrones:
 lines which points are at the same distance from the site of the wave origination. 
 By extracting intersection sites of isochrones of waves propagating from the source site to the destination site with isochrones of waves propagating from the destination site to the source site we can extract the shortest path. 
 Experiments report in  \cite{agladze1997finding} deal not with a maze but a space with two obstacles. The  approach would work in the maze as well.  The  BZ medium does not do any computation. 
 The results are obtained on a computer by analysing dynamics of the excitation wave fronts. 
Something is better than nothing: the approach is successfully used in unconventional robotics~\cite{adamatzky2004experimental, adamatzky2002collision}.

Another BZ based maze solver is proposed in \cite{rambidi2001chemical}: it exploits light-sensitive BZ reaction. Extraction of the path requires an extensive image analysis of the excitation dynamics, therefore this 
prototype is not worthy of discussion here.

\subsection{{\sc \large crystall}: Crystallization }
\label{crystallisation}
\index{crystallization}

In \cite{adamatzky1997cellular}  we proposed that a propagating excitation wave-front sets up pointers 
at each site of the medium. The pointers indicate a direction from where the wave-fronts came from. A shortest path towards the the source of the excitation can be recovered by following the pointers. This approach is experimentally implemented in \cite{adamatzky2009hot} where crystals play a role of pointers. 
A set of one-source-many-destinations shortest path in a room with obstacles is constructed 
using crystallisation of sodium acetate trihydrate~\cite{adamatzky2009hot}. The room with obstacles is technically a maze with irregularly shaped walls. Obstacles are made of an adhesive resin attached to a bottom of a Petri dish. A super-saturated clear solution of sodium acetate trihydrate is  poured into the Petri dish. 
The solution is cooled down.  A crystallisation is induced by briefly immersing an aluminium wire powdered with fine crystals of the sodium acetate into the target site in the solution. Crystals growing from nucleation sites bear distinctive elongated shapes expanding towards their proximal ends. Not only a crystal's overall shape but also the orientation of saw-tooth edges indicate the direction of the crystal's growth. The crystallisation patterns compete for space. A crystallisation pattern following longer than shortest path is unable to reach the source site because the space available is already occupied by the crystallisation pattern following the shortest path. Thus the direction of crystal growth can be detected by conventional image processing techniques, e.g. edge detection procedures, or by a complementary method of detecting directional uniformity of image domains. A configuration of local vectors, which indicate direction of crystallisation propagation, is calculated. A vector at each point indicates the direction from where the wave of crystallisation came from.

\section{Analysis}

With regards to time it takes the prototypes to solve the maze, i.e. to produce a traced path from the source to the 
destination, we can split the solvers into three groups:
\begin{itemize}
\item solve instantly (milliseconds or less): {\sc glow}, {\sc thermo}, {\sc vlsi}, {\sc crystall}
\item solve in minutes: {\sc assembly}, {\sc fluidic}, {\sc marangoni}, {\sc temperature}, {\sc wave}
\item solve in hours:  {\sc physarum i}, {\sc physarum ii}, {\sc epithelium}
\end{itemize}
The classification is very rough. We estimate the solution times based on reported sizes of mazes 
and expected propagation time of substances or perturbations in the maze solving substrates. 

Only {\sc crystall} visualises a gradient. Other prototypes do not. 
In  {\sc crystall} maze solver the temporal gradient can be seen by  an unaided eye as a pattern of crystal needles.

With regards to path tracing two groups can be specified:
\begin{itemize}
\item path is visible as a state of the solver, no computer is necessary: 
{\sc glow}, {\sc thermo}, {\sc assembly}, 
 {\sc fluidic},   {\sc marangoni},  {\sc temperature},    {\sc physarum I}, {\sc physarum II}, 
\item a computer is necessary to extract the path from the states of the solver: {\sc vlsi},  {\sc crystall},   {\sc wave},   {\sc epithelium}
\end{itemize}

In {\sc marangoni} no computer assistance is required if the path is traced using dyes. If the path is 
traced by a mobile droplet then we must record time lapse images of the travelling droplet and 
extract the path form the images. In {\sc epithelium}, if a single cells is used to trace the path 
then time lapse recording is necessary. However, if we send a procession of  epithelial cells, following one another, 
the path will become visible as a chain of cells. Because {\sc crystall} and  {\sc wave} definitely require an external computing device to visualise the path we exclude them from further analyse. We also exclude {\sc vlsi} because it a conventional hardware. 

Several maze solvers are `too demanding' in their requirements. Thus, {\sc glow} requires up 
to 0.1--500 Torr pressure and 2--30 kV electrical potential;  {\sc assembly} requires 1 kV electrical potential; 
 {\sc fluidic} requires 0.7 -- 2 Torr pressure, and  {\sc temperature} requires fluid to be scalding hot.
 
Let us check five remaining prototypes: 
{\sc thermo},   {\sc marangoni},    {\sc physarum~I},    {\sc physarum~II},   {\sc epithelium}.
 The prototypes {\sc marangoni} and {\sc epithelium} are not ideal because they 
 require some kind of a lab equipment not just a maze. The prototype {\sc physarum I} is computationally inefficient because it requires the slime mould to be placed in all channels of the maze. 

The remaining `contestants' are {\sc thermo} and {\sc physarum II}. Both prototypes are easy to make without a specialised lab equipment, and the whole setup of an experiment can be implemented in few minutes. 
It takes hours for {\sc physarum II} to trace and to visualise the path. 
The prototype {\sc thermo} calculates the path instantly.

To conclude, the maze solver {\sc thermo} is the fastest and the easiest to implement physical maze solver. 
However, the {\sc thermo} does not visualise the path. We need a thermal camera to see the path. 
The {\sc physarum II} visualises the path by morphology of the slime mould cell. However, the {\sc physarum II} `computes' the maze in many hours.

\section{Discussion}

\begin{figure}[!tbp]
\centering
\includegraphics[width=\textwidth]{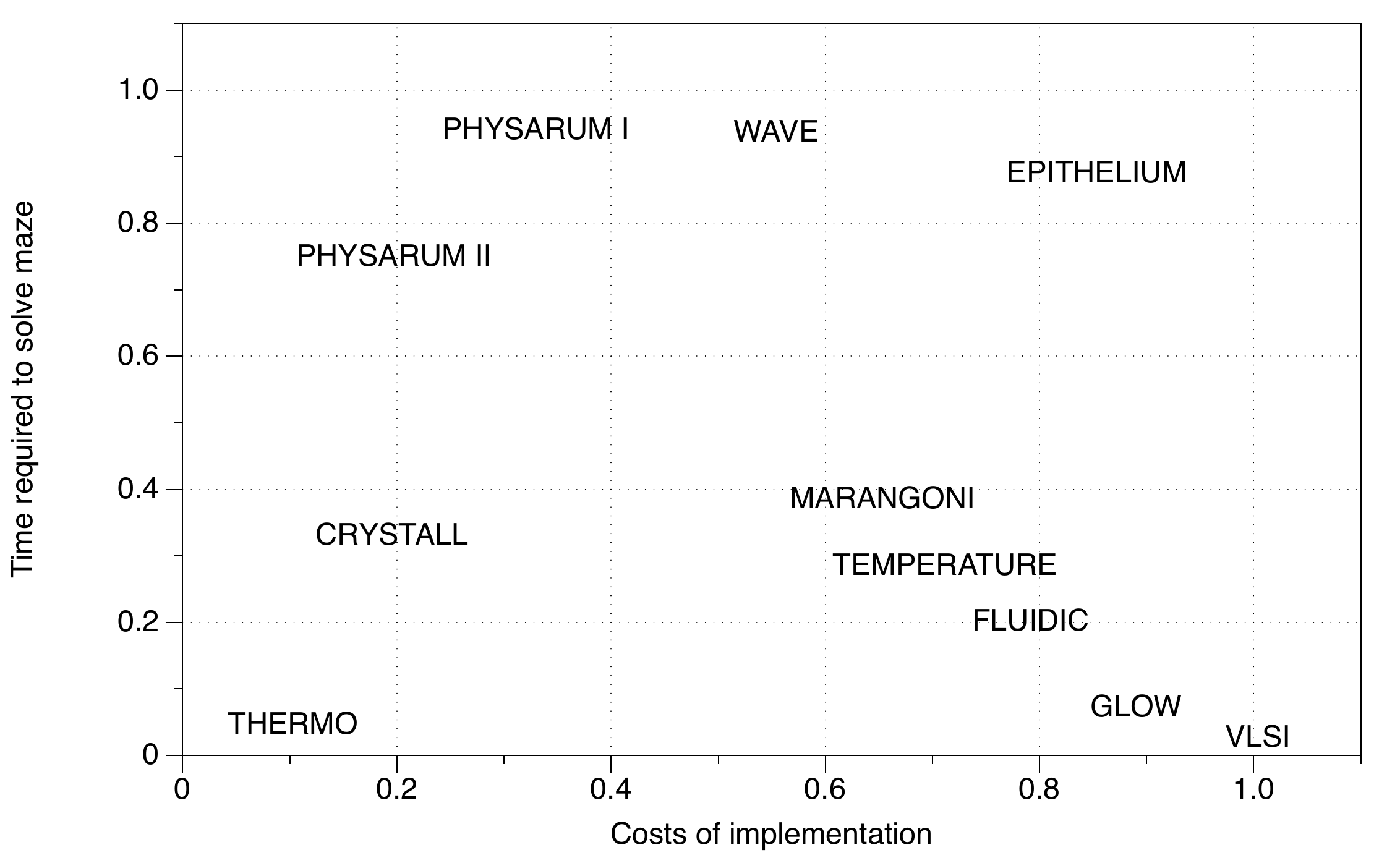}
\caption{Scatter plot of physical maze solvers based on estimated efforts of their prototyping and speed of maze solving.}
\label{speed_vs_costs}
\end{figure}

The experimental laboratory prototypes of maze solvers might look differently but they all implement the same 
two tasks. First, they explore mazes in parallel and develop resistance, chemical, thermal or temporal gradients. 
Thus they approximate a one-destination-many-sources set of shortest paths. Second, they trace a path from a given source site to the destination site using fluid flows, electricity or living cells.   Paths traced with electricity or flows are visualised with glow-charge gas, thermal camera, droplets, dyes. Paths traced with slime mould do not require any additional visualisation. Paths approximated with excitation wave-fronts or crystallisation patterns are visualised on computers.

A comparative plot of the maze solvers is shown in Fig.~\ref{speed_vs_costs}. We estimated speeds of maze solving from the experimental laboratory results reported, scaling all mazes to the same size. We estimated `costs of prototyping' based on descriptions of experimental setups.  The `costs' are not in monetary terms but include fuzzy estimates of efforts necessary to prepare a maze, experimental setup, auxiliary equipment, and an inconvenience of using extreme physical parameters. For example,  the glow-discharge maze solver requires  a high pressure of gas filling the maze and very high electrical potential; the maze solver exploiting temperature gradients demands filling fluid to be scalding hot. 

We have not discussed optical maze solving because we found reports only of theoretical designs~\cite{huang2004optical, okabayashi2011two, haistwave}. 
In \cite{huang2004optical} the symmetrical properties of optical systems are utilised, the light diffraction
is calculated and the path is extracted based on minimisation of the optical interferences. A detailed design of a  
two-dimensional nonlinear Fabry-Perot interferometer for maze solving is proposed in \cite{okabayashi2011two}. The optical solutions are interesting but impractical because they require a maze to be equipped with mirrors, so the light explores all paths in parallel;  and, every junction of the maze channels must be checked with interferometer. 
Also, to trace the path one must analyse all data on a computer. 

We have not included any results on solving mazes with living creatures but the slime mould.  Most papers published on experimental laboratory `maze solving' by living creatures do actually mean animals learning to make a left turn 
or a right turn in `T'- or`Y'-shaped junctions, as reported for ants~\cite{antmaze}, honey bees \cite{zhang2000maze}, nematodes~\cite{qin2007maze}.  Ants can approximate shortest paths~\cite{turner1907homing,goss1989self, goss1990trail, narendra2007homing}. Chances are high the ants can solve complex (not just `T'-junctions) mazes.   Some of the laboratory results, e.g. chemotactic behaviour and learning of nematodes \cite{reynolds2011chemotaxis,qin2007maze}, are enriched by simulations of a bit more complex than just `Y'-junction mazes.  Whatever case, most living creatures will not display the whole path in a maze. Slime mould does. In principle, plant roots can do as well. 

With regards to plants, apexes of plant roots show chemotactic behaviour. The roots can make a choice between direction of motion in Y-shape junctions and select a route towards the attractant~\cite{yokawa2014binary}. Experiments on plant root maze solving, where gravity is the only guiding force, were not successful  because the roots often stuck in a dead end~\cite{adamatzky2014towards}. This is not surprising thought cause the gravity does not provide a length-dependent tracing of a path. More experiments should be done in future with plant roots.

Do humans employ gradient developing and subsequent path tracing techniques when they solve maze visually? 
Given a full structure of a maze, i.e. being above the maze, humans solve the maze by scanning the maze,
memorising critical cues  and then tracing the path visually~\cite{zhao2013understanding}. The phases of the maze exploration and the path tracing are clearly distinct.  These phases can be seen as roughly corresponding to the gradient developing phase and  the path tracing, or visualisation, phase. 
There is a chance that a topological model of a maze is `physically' mapped onto a neuronal activity of a human brain cortex, where some neurons or their ensembles play a role of obstacles, other neurons are responsible for developing a gradient, and some neurons are responsible for tracing the path. In \cite{gourtzelidis2005mental}  a neuronal activity in superior parietal lobule of human subjects solving the maze is recorded with a whole-body magnetic resonance imaging and spectroscopy system. A mental traversing of the maze path is reflected in directional tuning of the volume unit. As \cite{gourtzelidis2005mental} suggest, this might indicate an existence of a directionally selective synaptic activity of spatially close neuronal ensembles. Micro-volumes of the parental lobe showing absence of tuning might be corresponding to obstacles of the maze.

\bibliographystyle{plain}
\bibliography{bibmaze}

\printindex

\end{document}